\documentclass[twocolumn,showpacs,preprintnumbers,amsmath,amssymb,showkeys,aps,prd,a4paper,byrevtex]{revtex4}

\usepackage[dvips]{graphicx}
\usepackage{dcolumn}
\usepackage{bm}
\usepackage{epsfig}
\usepackage{amsmath}
\usepackage{amsfonts}
\usepackage{amssymb,amscd}
\usepackage[latin1]{inputenc}
\usepackage{units}
\usepackage{hyperref}
\usepackage{braket}

\begin{document}

\newcommand{\qs}{Q_{\rm sat}}
\newcommand{\qsa}{Q_{\rm sat, A}}
\newcommand{\rr}{\mbox{\boldmath $r$}}
\newcommand{\rrn}{\mbox{$r$}} 
\newcommand{\pom}{I\!\!P} 
\newcommand{\rp}{\mbox{\boldmath $p$}} 
\newcommand{\rqq}{\mbox{\boldmath $q$}} 
\newcommand{\lsim}{\raisebox{-0.5mm}{$\stackrel{<}{\scriptstyle{\sim}}$}}
\newcommand{\gsim}{\raisebox{-0.5mm}{$\stackrel{>}{\scriptstyle{\sim}}$}}
\def\simge{\mathrel{%
   \rlap{\raise 0.511ex \hbox{$>$}}{\lower 0.511ex \hbox{$\sim$}}}}
\def\simle{\mathrel{
   \rlap{\raise 0.511ex \hbox{$<$}}{\lower 0.511ex \hbox{$\sim$}}}}

\title{Parton saturation scaling function for exclusive production of vector mesons and Deeply Virtual Compton Scattering}
\author{F.G. Ben$^{1}$, M.V.T. Machado$^{1}$ and W.K. Sauter$^{2}$}

\affiliation{$^{1}$High Energy Physics Phenomenology Group, GFPAE  IF-UFRGS \\
Caixa Postal 15051, CEP 91501-970, Porto Alegre, RS, Brazil\\
$^{2}$ Instituto de F\'{\i}sica e Matem\'atica,  Universidade
Federal de Pelotas\\
Caixa Postal 354, CEP 96010-900, Pelotas, RS, Brazil}

\date{\today}

\begin{abstract}
We provide a universal expression of cross sections for the exclusive vector meson production and Deeply Virtual Compton Scattering (DVCS) in photon-proton and photon-nucleus interactions based on the geometric scaling phenomenon. The theoretical parameterization based on the scaling property depends only  on the single variable $\tau_A = Q^2/Q_{\mathrm{sat}}^2$, where the saturation scale, $Q_{\mathrm{sat}}$, drives the energy dependence and the corresponding nuclear effects. This phenomenological result describes all available data from  DESY-HERA for $\rho,\,\phi,\,J/\psi$ production and DVCS measurements. A discussion is also carried out on the size of nuclear shadowing corrections on photon-nucleus interaction.
\end{abstract}
\maketitle

\section{Introduction \label{sec:intro}}

A striking property of the nonlinear  perturbative QCD  approaches for high energy deep inelastic electron-proton (or electron-nucleus) scattering (DIS)  is
the geometric scaling phenomenon. In the parton saturation based framework,  the total
$\gamma^* p$ and  $\gamma^* A$ cross sections are not a function of
the two independent variables $x$ (Bjorken scale) and $Q^2$ (photon virtuality), but is rather a
function \cite{travwaves} of a single scaling variable, $\tau_A = Q^2/Q_{\mathrm{sat,A}}^2$. Such a scaling is exact asymptotic solution of a general class of nonlinear evolution equations \cite{KPP,BK}  and it is a universal property of them. In particular, it corresponds to the traveling wave solutions of those equations. The saturation scale $Q_{\mathrm{sat,A}}^2(x;\,A)\propto xG_A(x,\, Q_{\mathrm{sat}}^2) /(\pi R_A^2)$, is connected with gluon saturation effects. At very small $x$, the strong rise of the gluon distribution function is expected to be controlled by saturation. It was demonstrated \cite{Iancu:2002tr}, however, that geometric scaling is not confined to the low momenta kinematic region, it is in fact preserved by the QCD evolution up to relative large virtualities. For proton target, it extends up to $Q^2 \simle Q_{\mathrm{sat}}^4 (x)/\Lambda^2_{\mathrm{QCD}}$,  provided one stays in small-$x$ region. For nuclear targets, that kinematic window is further enlarged due to the nuclear enhancement of the saturation scale, $Q_{\mathrm{sat,A}}^2\simeq A^{1/3}Q_{\mathrm{sat,p}}^2$. It was proven for the first time in Ref. \cite{Stasto:2000er} that the DESY-HERA $ep$ collider data
on the proton structure function $F_2$ present a scaling pattern 
at $x \leq 0.01$ and $Q^2 \leq \unit[400]{GeV^2}$. Similar behavior was further observed on  electron-nuclei processes \cite{Freund:2002ux} and on inclusive charm production \cite{magvicprl}. In Ref. \cite{MS} it was demonstrated that the data on diffractive DIS, $\gamma^*p\rightarrow Xp$, and other diffractive observables present geometric scaling on the variable $\tau_D = Q^2/Q_{\mathrm{sat}}^2(x_{\pom})$, in region $x_{\pom}<0.01$, where $x_{\pom} = (Q^2+M_X^2)/(Q^2+W^2)$. Moreover, the total cross sections for $\rho$, $\phi$ and $J/\psi$ are shown to present scaling on variable $\tau_V = (Q^2+M_V^2)/Q_{\mathrm{sat}}^2(x_{\pom})$. Nevertheless, \cite{MS} provides no theoretical or phenomenological expression for the scaling function. 

Concerning lepton-nucleus interactions, in Ref. \cite{Armesto_scal} the nuclear dependence of the $\gamma^*A$ cross section was absorbed in the $A$-dependence of the saturation scale via geometric scaling property. Namely, the $\gamma^*A$ cross section is obtained from the corresponding cross section for $\gamma^*p$ process in the form 
\begin{eqnarray}
 \sigma^{\gamma^*A}_{tot}\,(\tau_A)  =  \frac{\pi R_A^2}{\pi R_p^2}\,\sigma^{\gamma^*p}_{tot}\,\left(\tau_p\left[\frac{ \pi R_A^2}{A\pi R_p^2}\right]^{\Delta}\right),
\label{nuclear_scaling}
\end{eqnarray}
where $\tau_p = {Q^2}/{Q_{\mathrm{sat}}^2}$ is the saturation scale for a proton target.  The nuclear saturation scale was assumed to rise with the quotient of the transverse parton densities to the power $\Delta $. The nucleon saturation momentum is set to be $Q^2_{\mathrm{sat}}=\unit[(x_0/\bar{x})^{\lambda}]{GeV^2}$, where $x_0= 3.04\times 10^{-4}$, $\lambda=0.288$ and $\bar{x}=x\,[1+ (4m_f^2/Q^2)]$, with $m_f=\unit[0.14]{GeV}$, as taken from the usual Golec Biernat-W\"usthoff model \cite{GBW}. The nuclear radius is given by $R_A=(1.12 A^{1/3}-0.86 A^{-1/3})$ fm. The following scaling curve for the photoabsortion cross section was considered \cite{Armesto_scal}:
\begin{eqnarray}
  \sigma^{\gamma^* p}_{tot}\,(\tau_p) = \bar{\sigma}_0\,
  \left[ \gamma_E + \Gamma\left(0,\nu \right) +
         \ln \left(\nu \right) \right],
 \label{sigtot_param_tau}
\end{eqnarray}
where $\nu = a/\tau_p^{b}$, $\gamma_E$ is the Euler constant and $\Gamma\left(0,\nu \right)$
the incomplete Gamma function. The parameters for the proton case were obtained from a fit to the small-$x$ $ep$ DESY-HERA data, producing $a=1.868$, $b=0.746$ and the overall  normalization was fixed by $\bar\sigma_0=\unit[40.56]{\mu b}$. The parameters for the nuclear saturation scale were determined by fitting the available lepton-hadron data using the relation in Eq. (\ref{nuclear_scaling}) and the same scaling function, Eq. (\ref{sigtot_param_tau}). They obtained $\delta=1/\Delta = 0.79\pm0.02$ and $\pi R_p^2=\unit[1.55 \pm 0.02]{fm^2}$. 

In this work, we extend the approach presented in Ref. \cite{Armesto_scal}  to exclusive (diffractive) processes to describe also the observed scaling features demonstrated in Ref. \cite{MS}. Based on the eikonal model in impact parameter space, we provide an expression for the cross section for exclusive production of vector mesons and DVCS as well. This expression provides a reasonable description for the available data for $V=\rho,\,\phi,\, J/\psi$ and real photons. The results are improved  by allowing a global fit using the universal scaling expression which depends on very few parameters. These theoretical and phenomenological results have direct consequences on prediction for the future electron-ion colliders \cite{EICs} and also for vector meson photo-production measured in ultra-peripheral nucleus-nucleus collisions at the LHC \cite{Glauber1,Glauber2}. In the next section, we present the theoretical framework employed in the construction of the scaling function and analyse the data description discussing the possible limitations of approach and possible improvements. Finally, in the last section,  we present our main conclusions.

\section{Cross sections for exclusive vector meson production and DVCS \label{sec:sigma}}

The starting point in the derivation of our scaling formula for the exclusive cross section for the process, $\gamma^*h\rightarrow Eh$ (with $h=p,A$ and $E=V,\gamma$), is the eikonal model in the impact parameter space \cite{BC}. The elastic scattering amplitude $a(s,b)$ in general is assumed to be purely imaginary and the $s$-channel unitarity implies that $|a(s,b)|\leq 1$. In the eikonal approach, $a(s,b) = i(1-e^{-\Omega(s,b)})$, where the eikonal $\Omega$ is a real function. Thus, $P(s,b)=e^{-2\,\Omega(s,b)}$ gives the probability that no inelastic interaction takes place at impact parameter $b$. Assuming for simplicity a Gaussian form for the eikonal, $\Omega(s,b) = \nu(s)\exp\left( -b^2/R^2 \right)$, analytical expressions for total and elastic cross sections are generated,
\begin{eqnarray}
\sigma_{tot} & = & 2\int d^2b \,\mathrm{Im}\, a(s,b),\\
\sigma_{el} & = & \int d^2b \,|a(s,b)|^2.
\end{eqnarray}

Therefore, by use of the eikonal function in factorized form (with $\nu=\nu(s)$) discussed above one obtains,
\begin{eqnarray}
\label{sigtot}
\sigma_{tot} & = & 2\pi R^2\left[\ln(\nu) +\gamma_E+\Gamma\left(0,\nu \right) \right],\\
\sigma_{el} & = & \pi R^2\left[\ln \left(\frac{\nu}{2}\right) +\gamma_E-\Gamma\left(0,2\nu \right) + 2\,\Gamma\left(0,\nu \right)  \right].\label{sigel}
\end{eqnarray}

In hadronic models, the quantity $R$ depends on energy (in general, logarithmic behavior on energy). For the purpose presented here, the cross sections are being computed for fixed energy and thus we consider $R$ to be energy-independent. The Gaussian function is chosen as it allows  the $b$-integration to be analitically computed. Moreover, the two-dimensional Fourier transform of Gaussian profile has the exponential form, $d\sigma (\gamma^*p\rightarrow Ep)/dt\propto e^{-B_G|t|}$ (with $B_G \simeq  R^2$), which is supported by the data on exclusive production in DIS. More sofisticated models can be used, as the one corresponding to the power - like (dipole) form factor in momentum transfer representation \cite{GLM}, $S(b) = (\beta/\pi R^2)K_1(\beta)$ (with $\beta = \sqrt{8}b/R$).  It is clearly evident that the proposal of a scaling inclusive cross section having the form in Eq. (\ref{sigtot_param_tau}) relies on the total cross section from the eikonal model, Eq. (\ref{sigtot}), with the following identification, $\bar{\sigma}_0 = 2\pi R^2$ and $\nu = a/\tau_p^b$. The $a$ and $b$ parameters absorb the lost information when using a oversimplified photon wave-function overlap $\Phi^{\gamma^*\gamma^*}\propto \delta \left(r-1/Q\right)$ within the color dipole framework. Therefore, we will construct the scaling function for describing exclusive diffractive processes starting from Eq. (\ref{sigel}). The main point is to associate the exclusive vector meson production and DVCS process as a quasi-elastic scattering.

Before we proceed to the exclusive case, we would like to discuss in further detail the derivation of Eq. (\ref{sigtot_param_tau}) using the eikonal model. The starting point is to define the elementary dipole-target (proton)  scattering amplitude, excluding multiple scattering of the color dipole. Using color transparency and geometric scaling property one has, in general, for a fixed dipole size $r$  \cite{GBW}, $N_{q\bar{q}}(s,r) =\left( \frac{r^2Q_{\mathrm{sat}}^2}{4}\right)^{\gamma_s}$,  where effective $\gamma_s \simeq 1$ is the anomalous dimension. Now, we construct the elastic amplitude in $b$-space using the eikonal formalism (which includes the multiple dipole-target scattering)  and averaging over dipole sizes, 
\begin{eqnarray}
a(s,b) & = & \langle a(s,r,b) \rangle \equiv \int
  d^2r \int_0^1 dz \,\Phi^{\gamma^*\gamma^*}_{(T+L)}(z,r,Q^2) \nonumber \\&\times &  i\left[ 1-\exp\left(-N_{q\bar{q}}(s,r)S(b)\right) \right] ,\nonumber \\
&\approx &  i\left[ 1-\exp\left(-\frac{a\,S(b)}{\tau_p^{\gamma_s}}\right) \right],
\label{analitic}
\end{eqnarray}
with $\Phi^{\gamma^*\gamma^*} = \delta \left(r^2-\frac{A}{Q^2}\right) \delta\left(z-\frac{1}{2}\right)$, where $a = (4/A)^{-\gamma_s}$ and we can write $\nu= a/\tau_p^{b}$ (with $b=\gamma_s$). Using the recent determination of effective anomalous dimension $\gamma_s = 0.762 \pm 0.004$ \cite{Amir} and the typical values for $A=10$ from phenomenology \cite{McDermott}, we can estimate the parameters $a\simeq 2.01$ and $b\simeq 0.762$. They are quite close to the values $a=1.868$, $b=0.746$ found in Ref. \cite{Armesto_scal}.

For vector meson production, we have to include information related to the meson wave-function and in the DVCS case information on the real photon appearing in the final state. Adding this new information will modify the overall normalization in Eq. (\ref{sigel}) and possibly also the parameter $a$ and $b$ considered in Ref. \cite{Armesto_scal}. In order to clarify the situation, we shortly review the exclusive production within the color dipole framework. 

In an exclusive production process (vector mesons or DVCS) the photon splits into a dipole of transverse size $r$ and longitudinal momentum fraction $z$ which scatters elastically off the target (proton or nucleus), with virtuality $Q^2$ and recombines into a vector meson of mass $M_V$ or real photon of zero virtuality, $Q_{\gamma}=0$. Specifically for the former process, one introduces the wave-functions $\psi^{V,\lambda}_{f,h,\bar{h}}(z,r;M_V^2,Q^2)$ which describe the splitting of the vector meson with polarization $\lambda$ into the
dipole. An important ingredient to compute the production amplitude is the corresponding overlap function. These functions for the vector meson case and for DVCS are
\begin{eqnarray}
\label{overvm}
\Phi_{\lambda}^{\gamma^*V}(z,r;\mu^2)\!\!
 & = &\!\! \!\!\sum_{fh\bar{h}} \left[
   \psi^{V,\lambda}_{f,h,\bar{h}}(z,r;M_V^2)\right]^*
\psi^{\gamma^*,\lambda}_{f,h,\bar{h}}(z,r;Q^2),\\
\Phi^{\gamma^*\gamma}_T(z,r;Q^2) \!\!& = & \!\!\!\!\sum_{fh\bar{h}}
\left[\psi^{\gamma^*,T}_{f,h,\bar{h}}(z,r;0)\right]^*
\psi^{\gamma^*,T}_{f,h,\bar{h}}(z,r;Q^2),\label{overdv}
\end{eqnarray}
where the wavefunctions $\psi^{\gamma^*,\lambda}_{f,h,\bar{h}}(z,r;Q^2)$ describe the splitting of a virtual photon with polarization $\lambda\!=\!0,\pm 1$
into a dipole. The indices $h\!=\!\pm1$ and
$\bar{h}\!=\!\pm1$ denote the helicities of the quark and the
anti-quark composing the dipole of flavor $f$. Vector meson wave-functions rely on phenomenological models as the boosted Gaussian  (BG) \cite{mwfs1} and the light-cone Gaussian
(LCG) \cite{mwfs3}.

The overlap functions for exclusive processes are well known \cite{mwfs1,mwfs2,mwfs3} and we summarize them below. First, for the DVCS process one has,
\begin{eqnarray}
\label{ovdvcs}
\Phi^{\gamma^*\gamma}_T
   & = &\sum_f e_f^2 \frac{\alpha_e N_c}{2\pi^2}\left\{
           [z^2+(1-z)^2]\bar Q_f K_1(r\bar Q_f) m_f K_1(rm_f) \right. \nonumber \\
          & + & \left. m_f^2 K_0(r\bar Q_f) K_0(rm_f) \right\}, 
\end{eqnarray}
where $e_f$ and $m_f$ denote the charge and mass of the quark with flavor $f$ with $\bar Q_f^2 = z(1-z)Q^2+m_f^2$. Now, for the vector meson of polarizations $\lambda = L,T$  one obtains,
\begin{eqnarray}
\label{ovvml}
\Phi^{\gamma^*V}_L 
   & = & \hat{e}_f \sqrt{\frac{\alpha_e}{4\pi}} N_c \, 2QK_0(r\bar Q_f)  
\left[M_Vz(1-z)\phi_L(r,z) \right. \nonumber  \\
& +& \left. \delta\frac{m_f^2-\nabla_r^2}{M_V}\phi_L(r,z)\right],\\
\Phi^{\gamma^*V}_T 
   & = & \hat{e}_f \sqrt{\frac{\alpha_e}{4\pi}} N_c\left\{
             m_f^2 K_0(r\bar Q_f)\phi_T(r,z)  \right. \nonumber \\
           & - & \left.  [z^2+(1-z)^2]\bar Q_f K_1(r\bar Q_f) \partial_r\phi_T(r,z)  \right\},
\label{ovvmt}
\end{eqnarray}
where the constant $\hat{e}_f$ is an effective charge.  Those expressions are very similar to the photon ones except for the function $\phi_{\lambda}\propto f_{\lambda}(z,M_V)\exp\left[-r^2/(2R_{\lambda}^2)\right]$ which it is related to the
vertex function and depends on the model.

Accordingly, considering the scattering amplitude for the exclusive process, $\gamma^*p\rightarrow Ep$ ($E=V,\gamma$), pure imaginary and disregarding real part contribution and skewness corrections as well, the differential cross-section reads 
\begin{eqnarray}
\frac{d\sigma^{\gamma^*p\rightarrow Ep}}{dt} & = & 
  \frac{1}{16\pi} \left|\int d^2b\, \int
  d^2r \int_0^1 dz \left(\Phi_{T}^{\gamma^*E}+\Phi_{L}^{\gamma^*E}\right) \right. \nonumber \\
 &\times & \left.  \exp\left[i\mathbf{q}\cdot\left(\mathbf{b}-z\mathbf{r}\right)\right]\,a (r,b,Y)\right|^2,\label{eq:sigmaeldip} 
\end{eqnarray}
where $a(r,b,Y)$ is the dipole-target scattering amplitude
and carries all the energy dependence via the rapidity $Y$ which is
obtained from the center-of-mass energy $W$ and the  typical momentum scale for the exclusive process. For instance, for vector meson production of mass $M_V$ one writes $Y = \log[(W^2+Q^2)/(M_V^2+Q^2)]$. Moreover,  one has $t=-\mathbf{q}^2$, where  $\mathbf{q}$ represents the transverse momentum transfered by the target during the collision.

From Eqs. (\ref{ovvml}) and (\ref{ovvmt}), the main features about the meson properties are embedded into the $\phi_{\lambda}$ function. In general, the wave-functions in the mixed representation $(z,\mathbf{r})$ are obtained from the momentum representation $(z,\mathbf{k}_{\perp})$ wave-functions using a Fourier transform,
\begin{equation}
\label{fouriertrans}
\phi_V(r,z) = \int \frac{d^{2}k_{\perp}}{ 4\pi^2 }\, \phi_{V}(z,k_{\perp})\,
e^{i\mathbf{r}\cdot \mathbf{k}_{\perp}}.
\end{equation}

In the simplest case one considers that a heavy $q$ and $\bar{q}$ have the same longitudinal momentum fraction and that the transverse momentum is quite small. Such an hypothesis yields $\phi_{V}(z,k_t)=N_V\ \delta(z-1/2)\,\delta^2(k_{\perp})$. The only free parameter is the normalization, $N_V$, which can be determined by fixing the partial width for $V \rightarrow e^+e^-$ to the
experimentally measured value,
\begin{equation}
\Gamma^{V}_{e^+e^-}=\frac{32 \pi\alpha_e^2 e_q^2}{M_V}\left| \int dz \int \frac{
d^{2}k_{\perp}}{8\pi^{3/2}} \phi_V (z,k_{\perp}) \right|^2.
\end{equation}

\begin{table}[t]
\begin{tabular}{||c|c|c|c||} \hline\hline
Meson & $M_V$ (GeV) & $f_V$ (GeV) & $\hat{e}_V$  \\ \hline
$J/\psi$ & 3.097 & 0.274 & 2/3 \\
$\phi$ & 1.019 & 0.076 & 1/3 \\
$\rho$ & 0.776 & 0.156 & 1/$\sqrt{2}$ \\ \hline\hline
\end{tabular}
\caption{\label{tab:par} Values of the parameters $M_V$, $f_V$ and $\hat{e}_V$ from Ref. \cite{Kowalski:2006hc}.}
\end{table}

Therefore, the wave-function in the mixed representation obtained via
Eq. (\ref{fouriertrans}) is written as
\begin{eqnarray}
\phi_V(r,z) & = & \frac{1}{2 M_V} \frac{\sqrt{\pi}}{2\sqrt{6}\, e_q\,
\alpha_e}\sqrt{\frac{3\Gamma^{V}_{e^+e^-} M_V}{\pi}} \,\delta \left(z-\frac{1}{2}\right),\nonumber \\
& = & \frac{1}{2 M_V} \frac{\sqrt{\pi}}{\sqrt{6}\, e_q} f_V\,\delta \left(z-\frac{1}{2}\right),
\label{wfapprox}
\end{eqnarray}
where $3\,\Gamma^{V}_{e^+e^-} M_V = 4\pi \alpha_e^2f_V^2$, with $f_V$ being the coupling of the meson to the electromagnetic current. We will use the approximation in Eq. (\ref{wfapprox}) in the following discussion. The wavefunction overlap appearing in Eq. (\ref{eq:sigmaeldip}) takes the simplified form, $\Phi_{(T+L)}^{\gamma^*V} \propto \delta (r^2-r_Q^2)\delta (z-1/2)$, where $r_Q^2 = A_Q/(Q^2+m_V^2)$. That is, the exclusive production of vector mesons, $\gamma^*p\rightarrow Vp$, in deep inelastic scattering is a hard scattering process in which the transverse size $r_Q$ of quark configurations that dominate the production amplitude are under theoretical control.  The quantity $A_Q$ is now process dependent (distinct for light and heavy mesons) \cite{KNNZ} and for a naive estimation one can use an average value $A_Q=4$ \cite{KNNZ}. Repeting the discussion after Eq. (\ref{analitic}), for exclusive production of vector mesons one has $a=(4/A_Q)^{-\gamma_s}\approx 1$. In our analysis we allow the parameters $a$ (and $b$) to be  process dependent. We have shown in equation Eq. (\ref{analitic}) that the parameters $a$ and $b$ are correlated. Moreover, the parameter $a$ is connected to the peak (in $r$ variable) of the overlap function, which is process dependent. Indeed,  it depends also in the corresponding kinematics as already presented in Ref. \cite{KNNZ}.  In exact geometric scaling models, the parameter $b=\gamma_s$ (the effective anomalous dimension) should be process independent. Namely, it is a constant between 0.5 (BFKL dynamics) and  1 (DGLAP dynamics). However, in extended geometric scaling models (see, for instance Ref. \cite{Iancu:2002tr}) it acquires a $r$-dependence and this can be translated in a change on $b$ value depending on (specific wavefunctions overlap) the process.

The exclusive processes described above can be directly compared to the inclusive case, in which the total absorption cross section is related to the imaginary part of scattering amplitude,
\begin{eqnarray}
\sigma^{\gamma^*p\rightarrow X} & = & \int
  d^2r \int_0^1 dz \left[\Phi^{\gamma^*\gamma^*}_T(z,r,Q^2) + \Phi^{\gamma^*\gamma^*}_L(z,r,Q^2) \right] \nonumber \\
&\times &  2 \int d^2b \,a(r,b,Y) \equiv 2\int d^2b\,\mathrm{Im}\left\langle a(Y,b) \right\rangle ,
\label{eq:sigmaabsdip} 
\end{eqnarray}
where the quantum mechanics average is taken over the initial and final state virtual photons,
\begin{eqnarray}
 \Phi^{\gamma^*\gamma^*}_T+\Phi^{\gamma^*\gamma^*}_L
    &=& \sum_f e_f^2 \frac{\alpha_e N_c}{2\pi^2}\left \{[z^2+(1-z)^2]\bar Q_f^2 K_1^2(r\bar Q_f) \right. \nonumber \\
&+ & \left.\left[ 4 Q^2 z^2(1-z)^2 + m_f^2\right] K_0^2(r\bar 
Q_f) \right \}.
\end{eqnarray}

Finally, we can write down a scaling curve for exclusive processes using the eikonal model result in Eq. (\ref{sigel}) and the geometric scaling assumption as derived in Eq. (\ref{analitic}).  The remaining ingredient it to redefine the overall normalization for the different cases. The total cross section for an exclusive process (DVCS and vector mesons) is written  as follows, 
\begin{eqnarray}
\sigma (\gamma^*p\rightarrow Ep)  =  \frac{\bar{\sigma}_E}{2}\left[\ln \left(\frac{\nu}{2}\right) +\gamma_E-\Gamma\left(0,2\nu \right) + 2\,\Gamma\left(0,\nu \right)  \right],\label{sigexcl}
\end{eqnarray}
where $\bar{\sigma}_E = \bar{\sigma}_V$ in case of vector mesons and $\bar{\sigma}_E = \bar{\sigma}_{\mathrm{DVCS}}$  for DVCS process. In both cases, $\nu = a/\tau^b$, with $\tau = (Q^2 + M_V^2)/Q^2_\mathrm{sat}$ for exclusive production of mesons and $\tau = Q^2/Q^2_\mathrm{sat}$ for DVCS. Explicitly, the overall normalization of cross sections is obtained from the inspection of the overlap functions in Eqs. (\ref{eq:sigmaeldip}) and (\ref{eq:sigmaabsdip}). Therefore, the final expressions for the overall normalization in our scaling function are given by
\begin{eqnarray}
\label{sig0dvcs}
\bar{\sigma}_{\mathrm{DVCS}} & = & \left(\alpha_e\,\sum_f e_f^2  \right)\bar{\sigma}_0, \\
\label{sig0v}
\bar{\sigma}_V & = & \frac{4\pi \hat{e}^2_ff_V^2}{M_V^2\left(\sum_f e_f^2 \right)}\bar{\sigma}_0.
\end{eqnarray}

In next section we wiil test the assumption above and discuss the consequences for the case where nuclei targets are considered. The stability and model dependence for the parameterss ($a$ and $b$) are analysed.

\begin{table}[t]
\begin{tabular}{||c|c|c|c|c||} \hline\hline
 & $a$ & $b$ & $\bar{\sigma}_0$($\mu$b) & $\chi^2/\mathrm{d.o.f.}$  \\ \hline
DVCS & & & & \\ \hline
ASW & 1.868 & 0.746 & 40.56 & 3.248 \\ \hline
Fit 1 & 1.313 & 0.769 & 114.610 & 0.768 \\ \hline
Fit 2 & 1.938 & 0.710 & 40.56 & 0.754 \\ \hline
$J/\psi$ & & & & \\ \hline
ASW & 1.868 & 0.746 & 40.56 & 4.567 \\ \hline
Fit 1 & 1.851 & 0.733 & 52.524 & 1.083 \\ \hline
Fit 2 & 1.919 & 0.704 & 40.56 & 1.183 \\ \hline
$\phi$ & & & & \\ \hline
ASW & 1.868 & 0.746 & 40.56 & 21.706 \\ \hline
Fit 1 & 1.936 & 0.750 & 72.717 & 8.843 \\ \hline
Fit 2 & 2.061 & 0.695 & 40.56 & 14.419 \\ \hline
$\rho$ & & & & \\ \hline
ASW & 1.868 & 0.746 & 40.56 & 529.004 \\ \hline
Fit 1 & 1.684 & 0.916 & 27.333 & 1.266 \\ \hline
Fit 2 & 1.467 & 0.943 & 40.56 & 1.011 \\ \hline \hline
\end{tabular}
\caption{\label{tab:chi} Summary of fitting procedure. ASW is the result using the original parameters from the fit to $ep$ HERA data \cite{Armesto_scal}. Fit 1 adjusts parameters $a$, $b$ and normalization $\bar{\sigma}_0$. Fit 2 adjusts $a$ and $b$ keeping fixed $\bar\sigma_0=\unit[40.56]{\mu b}$ (as for the inclusive case).}
\end{table}

\section{Results \label{sec:result}}
\begin{figure*}[t]
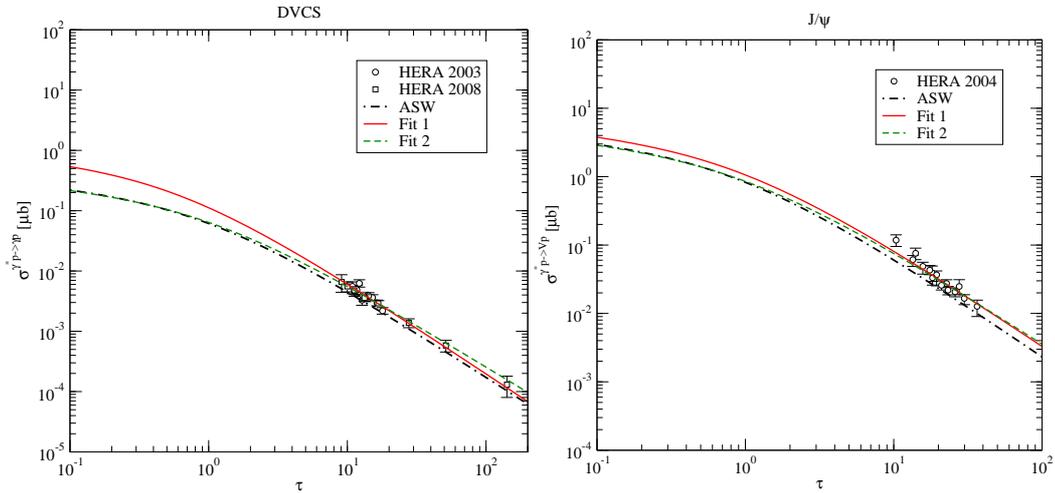

 \includegraphics[width=0.8\columnwidth]{dvcs.eps}\includegraphics[width=0.8\columnwidth]{jpsi.eps}
 \caption{The cross section for DVCS (left panel) \cite{dvcs} and $J/\psi$ production (right panel) \cite{jpsi} as a function of the corresponding scaling variable $\tau$. The ASW result is represented by a dot-dashed line, the Fit 1 by solid lines and Fit 2 by the dashed ones.} 
 \label{fig:1}
 \end{figure*}

\begin{figure*}[t]
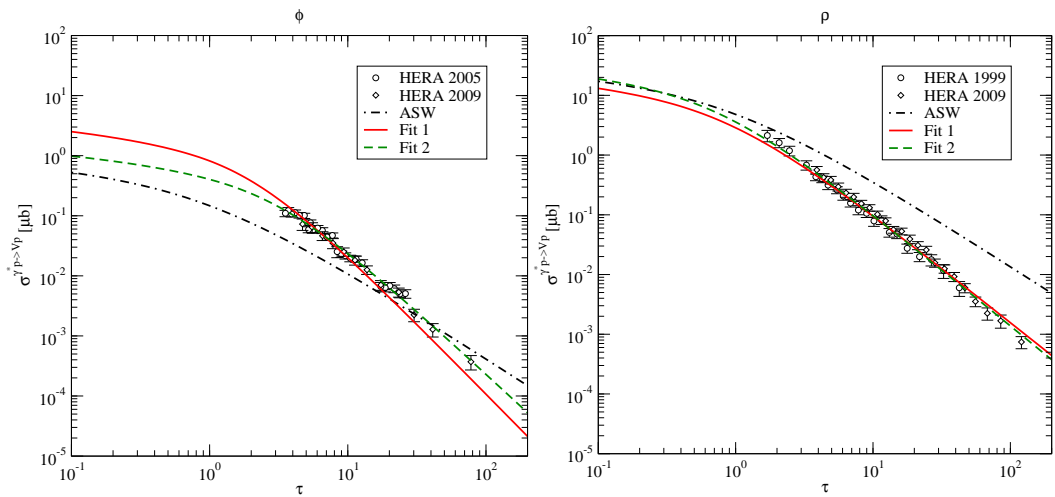

\includegraphics[width=0.8\columnwidth]{phi.eps}\includegraphics[width=0.8\columnwidth]{rho.eps}
\caption{The cross section for $\phi$ \cite{phi} (left panel) and $\rho$ \cite{rho} production (right panel) as a function of the corresponding scaling variable $\tau$. The ASW result is represented by a dot-dashed line, the Fit 1 by solid lines and Fit 2 by the dashed ones.} 
\label{fig:2}
\end{figure*}

Let us now compare the scaling curve, Eq. (\ref{sigexcl}), to the available experimental data in small-$x$ lepton-proton collisions. The data sets we have considered are presented in Refs. \cite{dvcs,rho,phi,jpsi}. The values of parameters $M_V$, $f_V$ and $\hat{e}_V$ were taken  from Ref. \cite{Kowalski:2006hc} and displayed in Tab. \ref{tab:par}. We perform a fit to the experimental data using MINPACK routines~\cite{minpack} for choices of sets of parameters, described in the following. Our results are presented in Table \ref{tab:chi} and in the Figures \ref{fig:1} and \ref{fig:2} as a function of the scaling variable $\tau$. Explicitly, the scaling variable is $\tau = \tau_V = (Q^2 + M_V^2)/Q^2_\mathrm{sat}(x)$ for exclusive production of mesons and $\tau = Q^2/Q^2_\mathrm{sat}(x)$ for DVCS, with $Q^2_{\mathrm{sat}}(x)=\unit[(x_0/\bar{x})^{\lambda}]{GeV^2}$ as discussed in the introduction section.

We use two different choices to perform the fits. The first one, labeled ``Fit 1" in the figures and table, adjusts all the three parameters ($a$, $b$ and $\bar{\sigma}_0$). The other one, labeled ``Fit 2" in the figures, fits $a$, $b$ parameters, maintaining fixed $\bar{\sigma}_0 = \unit[40.56]{\mu b}$. In general both fits describe in good agreement the available data for all observables (with the exception of $\phi$ meson) for photon-proton interactions. It is very clear that the quality of fit for Fit 1 and Fit 2 are somewhat equivalent. Fit 2 is a straightforward extension of the celebrated scaling curve presented in Ref. \cite{Armesto_scal} for the inclusive case. The overall normalization $\bar{\sigma}_0$ is common to inclusive and exclusive photon-target processes. For the sake of completeness, we also include the result using the original values for the parameters from the fitting to inclusive data \cite{Armesto_scal} (labeled by ASW in the curves) .

In Fig.~\ref{fig:1} the cross section for DVCS and $J/\psi$ production are presented as a function of the scaling variable $\tau$.   Our scaling curves are represented by the solid (Fit 1) and dashed (Fit 2) lines in the figure. From Table \ref{tab:chi}, the quality of fit is very good even for Fit 2 which considers only two parameters and $\chi^2/\mathrm{d.o.f.}\simeq 1$. It is interesting to note the stability of parameters $a$ and $b$ in both cases and the proximity with the values obtained for the inclusive case $a=1.868$ and $b=0.746$  \cite{Armesto_scal} (deviation of $\approx 5 $\%). 

In Fig. \ref{fig:2} the scaling data for $\rho$ \cite{rho} and $\phi$ \cite{phi} meson are presented, using the same notation as in the previous figure.  The corresponding $\chi^2/\mathrm{dof}$ for the $\phi$ case is problematic even including the recent data from DESY-HERA. The origin of such a discrepancy should be treated in a deeper study as the $\phi$ meson is in the transition region from light to heavy mesons. On the other hand, the quality of fit for the $\rho$  case is excellent where now $\chi^2/\mathrm{d.o.f.}\simeq 1$ in Fit 2. We verify also that the parameters $a$ and $b$ deviate more strongly from the original values for the inclusive case. We will discuss the consequences of these facts in the following. The main explanation on why the  fit with three parameters
(Fit 1) provides almost always worse $\chi^2/\mathrm{dof}$ than the one with two parameters (Fit 2) is the complex bahavior of the scaling curve as a function of the original seed in the fitting procedure. It probably means that the minimisation procedure went to a local minimum (clearly, Fit 2 with the parameter $\bar{\sigma}_0=40.56$ $\mu$b gives a smaller $\chi^2$).  A possible improvement in our analysis would be to consider a global analysis for exclusive observables. In any case, Fit 2 still gives a reasonable data description (we have tested the fit using the option where $a$ and $b$ are very larger than the original ASW parameters and obtain $a=6.175$ and $b=1.07$ with $\chi^2/\mathrm{d.o.f.}\simeq 1.294$). We clearly have a problem to describe the $\phi$ case, and the coherence could be improved by including also a scale dependence of the t slope. It is well known that $t$-slope strongly depends on $Q^2$ for $\rho$ and $\phi$, while flat for DVCS and $J/\psi$. For instance, parametrizing $B_V(Q^2)\approx R(Q^2)$ and introducing it in to the fit could improve the coherence of parameters among the different processes. Here, the cross section (integrated on t) was obtained using the simplified gaussian behavior for the impact parameter dependence. The differential cross section was not calculated and probably would present distinct behavior when comparing DVCS, $J/\psi$  and light mesons.

The main features from the scaling curve can be traced from the asymptotic limit of Eq. (\ref{sigexcl}). For instance, for $\nu\gg 1$ which corresponds to large values of scaling variable $\tau$ one has $\sigma (\gamma^*p\rightarrow Ep)\approx (\bar{\sigma}_E/2)\left[\ln(\nu/2)+\gamma_E \right] \propto -b \ln\, (\tau)$.  On the other hand, in the limit $\nu\ll 1$ related to the large $\tau$ values one has  $\sigma (\gamma^*p\rightarrow Ep)\propto \tau^{-b}$. This discussion is important in the photoproduction case, where one has the smallest values for the scaling variable $\tau $. This is the case for the  few available data points for $\omega $, $\psi (2S)$ and $\Upsilon$ states \cite{Capua}, mostly of them measured for $Q^2=0$ and having large experimental uncertainties.  For light mesons at $W = 200$ GeV (HERA-HERA), one can use the asymptotic expression for $\nu\gg 1$ and an approximation in the photoproduction limit would be $\sigma (\gamma p \rightarrow VP) \approx (\bar{\sigma}_V/2)(1+\gamma_E)$  modulo logarithmic corrections. It gives $1.38\times 10^{-1}$ $\mu$b for $\omega$, which it is consistent with order of magnitude of data. For $\psi (2S)$ and $\Upsilon$, the $\tau_V$ variable is still  large and then we can obtain the following approximation, $\sigma (\gamma p \rightarrow VP) \approx (\bar{\sigma}_V/2) \tau_V^{-b}$ (using $b=0.746$). This produces 
$3.3\times 10^{-4}$ $\mu$b for $\Upsilon$ and $4\times 10^{-2}$ $\mu$b for $\psi(2S)$.  In any case, these crude estimations are consistent with the 
current experimental results \cite{Capua}.

The geometric scaling present in the lepton-proton cross sections for exclusive processes, as quantified by Eq. (\ref{sigexcl}),  is translated to the scattering on nuclear targets at high energies. Following the same arguments given in Ref. \cite{Armesto_scal}, the atomic number dependence is absorbed in the nuclear saturation scale and on the overall normalization related to the nuclear radius. Therefore, the cross section for lepton-nuclei scattering takes the following form, 
\begin{eqnarray}
 \sigma^{\gamma^*A\rightarrow EA}\,(\tau_A)  =  \frac{\pi R_A^2}{\pi R_p^2}\,\sigma^{\gamma^*p\rightarrow Ep}\,\left(\tau = \tau_A\right),
\label{nuclear_scaling2}
\end{eqnarray}
where the scaling variable in nuclear case is $\tau_A = \tau_p[\pi R_A^2/(A 
\pi R_p^2)]^{\Delta}$. In particular, we expect that for large $\tau_A$ the relation  is $\sigma (\gamma^*A\rightarrow EA)\propto R_A^2\,\tau_A^{-b}=R_A^2 \tau_p^{-b}(A^{1/3})^{\frac{b}{\delta}}$.  As the current data on nuclear targets are quite scarce at small-$x$ region, the scaling formula above can be tested in future measurements in EICs or in ultraperipheral heavy ions collisions. The robustness of the geometric scaling treatment for the interaction is quite impressive and similar scaling properties have been proved theoretically and experimentaly, for instance in charged hadron production \cite{prazlach} and in prompt photon production \cite{prazlapf} on $pA$ and $AA$ collisions in colliders energy regime.

\begin{figure*}[t]
\includegraphics[width=0.8\columnwidth]{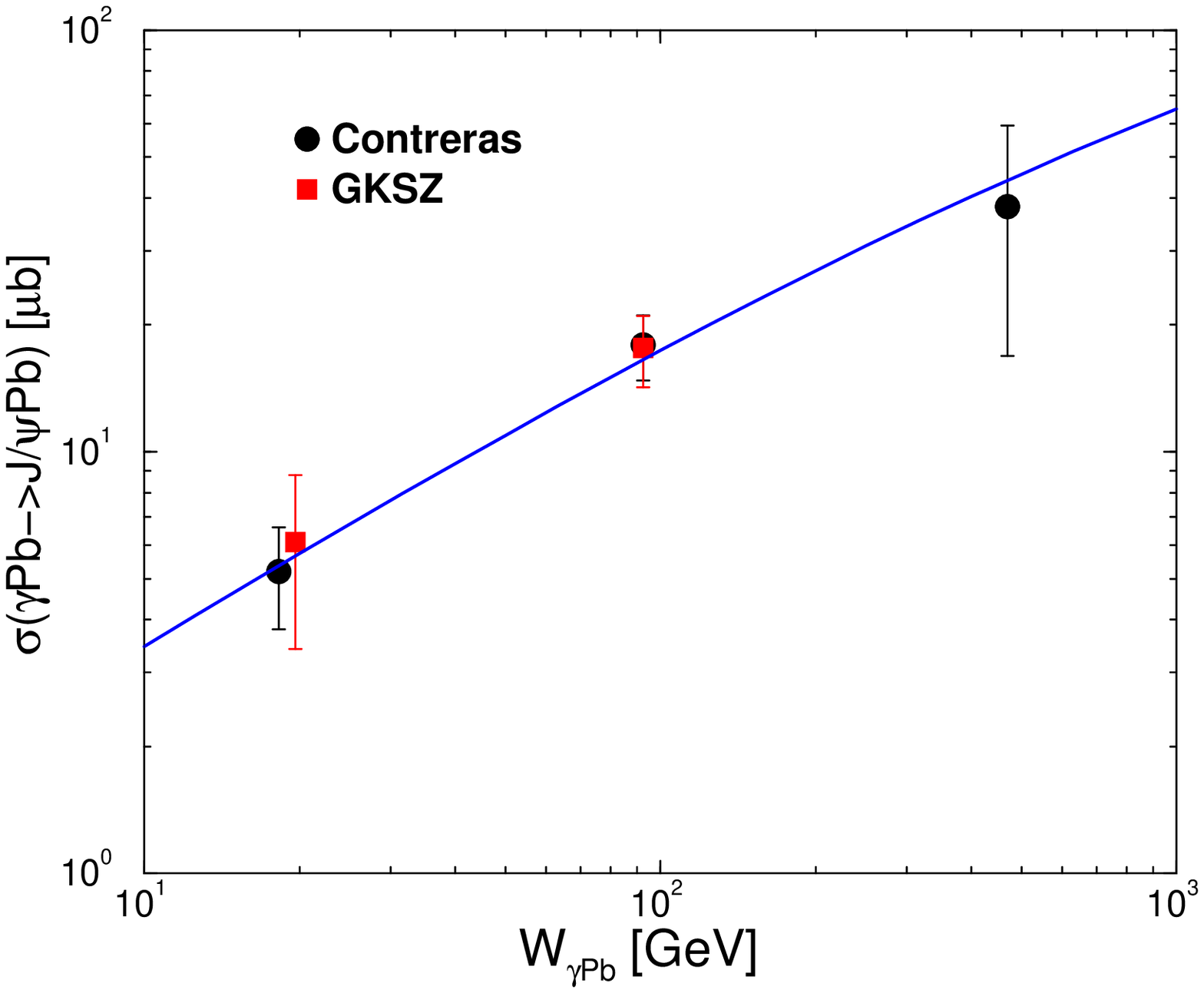}\includegraphics[width=0.8\columnwidth]{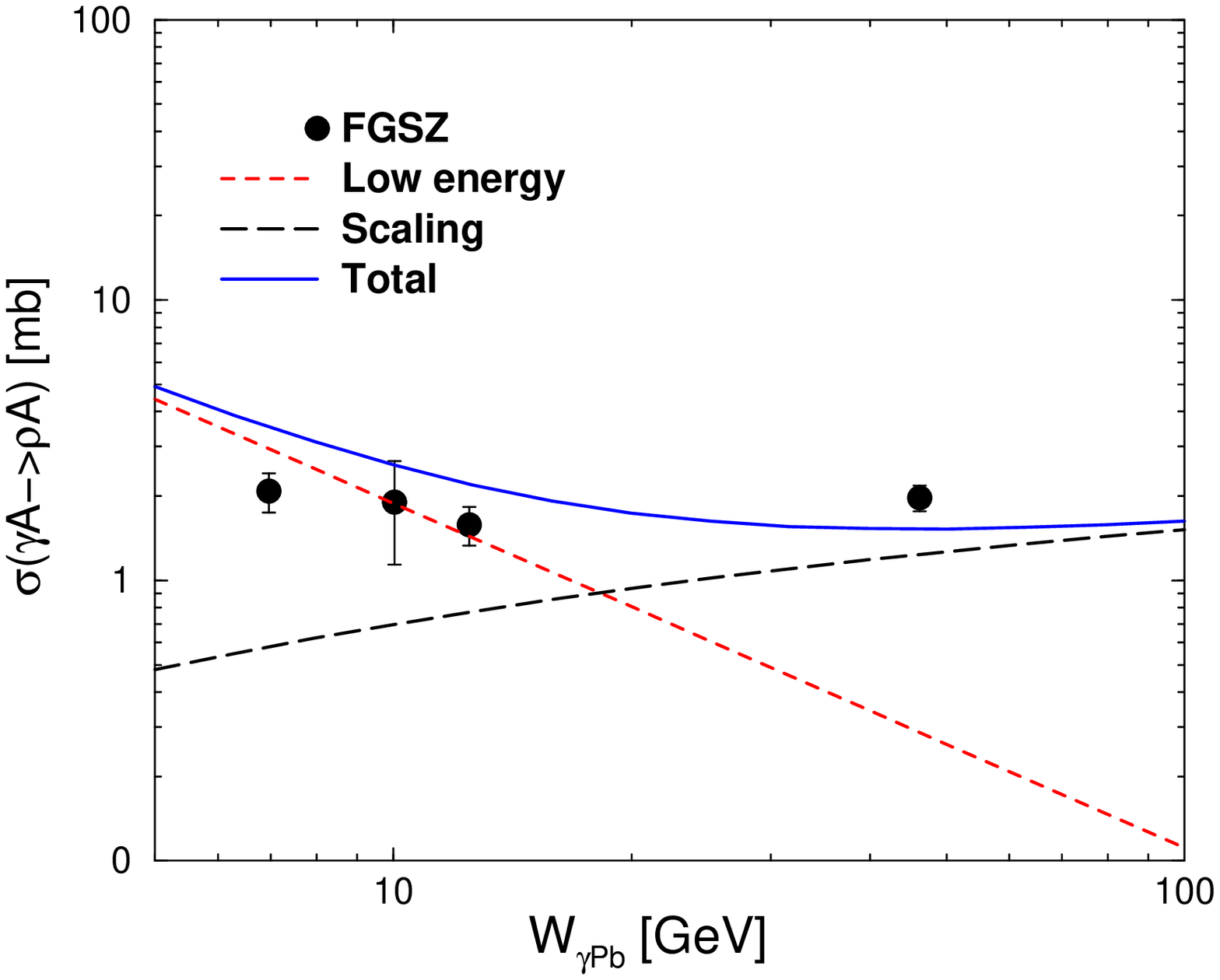}
\caption{The cross section $\sigma (\gamma A\rightarrow V A)$ for $V = J/\psi$  (left panel) and $V=\rho$  production (right panel) as a function of the corresponding photon-nucleus energy (see discussion in text).} 
\label{fig:3}
\end{figure*}

Still discussing the nuclear case, the fitted values for the $b$ parameter have strong consequences on the role played by the nuclear shadowing for the distinct final states we have considered here. As defined in \cite{Armesto_scal}, the relation between the nuclear saturation scale, $Q_{\mathrm{sat},A}$, and the proton one, $Q_{\mathrm{sat},p}$, is given by,
\begin{eqnarray}
Q_{\mathrm{sat},A}^2 = Q_{\mathrm{sat},p}^2\left[\frac{A\pi R_p^2}{ \pi R_A^2}\right]^{\Delta}, \,\,\delta = \frac{1}{\Delta} = 0.79\pm 0.02,
\end{eqnarray}
which implies that the small-$x$ data on nuclear structure functions $F_2^A$ favour an enhancement of the nuclear saturation scale faster than the usual $Q_{\mathrm{sat},A}^2 = A^{1/3}Q_{\mathrm{sat},p}^2$ estimation. Moreover, the coincidence of the $b$ and $\delta$ parameters in the inclusive case indicates absence of shadowing in the nuclear parton distributions at $Q^2\gg Q_{\mathrm{sat},A}^2$. We see that the situation is  different in the exclusive case. For instance, for DVCS at high $Q^2$ and for electroproduction of $J/\psi$ the parameter $b$ is still similar to the inclusive case. On the contrary, for the light $\rho$ meson the deviation is quite large, where $b/\delta >1$. This indicates that a study  for the value of $\delta$ parameter  in a electron-ion collider is quite important. Probably, its value for exclusive processes in $\gamma^*A$ interactions should be larger than for the inclusive case allowing for a strong nuclear shadowing even for milder values of photon virtualities. In order to qualify this discussion, we compare our predictions to the cross sections, $\sigma (\gamma A\rightarrow V A)$, extracted from the ultraperipheral $AA$ data from RHIC and LHC. In Fig. \ref{fig:3}-a is shown the photonuclear cross section for $J/\psi$ production using the parameters of Fit 2 as a function of photon-nucleus energy, $W_{\gamma A}$. The extracted cross sections are from Ref. \cite{Guzey:2013xba}  (labeled GKSZ) and Ref. \cite{Contreras:2016pkc}   (labeled Contreras) and the data description is quite reasonable. A similar analysis could be done also for $\psi (2S)$ state using an extracted cross section as discussed for instance in Ref. \cite{Guzey:2016piu}. In Fig. \ref{fig:3}-b, the prediction for $\rho$ production is considered also using the parameters from Fit 2. The situation here is more complicated as the extracted cross sections include the UPC data from RHIC (AuAu collisions) which correspond to low energy range. The cross section at higher energy is obtained from the LHC PbPb data. We have considered the cross sections values available in Ref. \cite{Frankfurt:2015cwa} (labeled FGSZ). For simplicity, at low energy we consider a black disk scaling  following Ref. \cite{Klein:1999qj}, i.e. $\sigma(\gamma A \rightarrow  \rho A)\simeq A^{4/3}\sigma(\gamma p \rightarrow  \rho p)=A^{4/3}YW^{-\eta}$ (with $Y= 26$ $\mu$b and $\eta = 1.23$). The low energy contribution corresponds to the dashed curve, the geometric scaling prediction is the dot-dashed curve and the total result is represented by the solid curve. The data description is still reasonable given the simplicity of the approach, with the total result underestimating the high energy extracted cross section.

\section{Summary and conclusions \label{sec:sumcon}} 

This work demonstrates that by assuming geometric scaling phenomenon in exclusive processes at small-$x$ and simple considerations on the scope of  eikonal model, one is able to describe the available data on DVCS and vector meson production on nucleon target with a universal scaling function without any further parameter. We establish that the geometric scaling parametrization can be extrapolated to nuclear targets to be tested in future EICs or in ultra-peripheral collisions.  This implies that such dimensionless scale absorbs their energy and atomic number dependences. The scaling curve is derived for the first time for exclusive case, generalizing the scaling curve found for the inclusive cace. The identification of the physical meaning of the corresponding parameters is done and the implications of those values in the nuclear case has been discussed.The application of the current result for the diffractive structure function and the ratio $\sigma_D/\sigma_{tot}$ is straightforward. 

\begin{acknowledgments}


This work was financed by the Brazilian funding agency CNPq. The authors are grateful to Laurent Favart for helpfull discussions and comments.

\end{acknowledgments}

\end{document}